\documentclass[astronomy,article,submit,oneauthor]{Definitions/mdpi} 
\usepackage{aas_macros}

\firstpage{1} 
\pubvolume{1}
\issuenum{1}
\articlenumber{0}
\pubyear{2025}
\copyrightyear{2025}
\datereceived{ } 
\daterevised{ } % Comment out if no revised date
\dateaccepted{\today} 
\datepublished{ } 
\hreflink{https://doi.org/} % If needed use \linebreak
\Title{Deciphering the Electron Spectral Hardening in AMS-02}

\TitleCitation{Deciphering the Electron Spectral Hardening in AMS-02}

\Author{Carmelo Evoli $^{1,2}$\orcidA{}}

\AuthorNames{Carmelo Evoli}
\AuthorCitation{Evoli, C.}

\address{%
$^{1}$ \quad Gran Sasso Science Institute (GSSI), Viale Francesco Crispi 7, 67100 L’Aquila, Italy\\
$^{2}$ \quad INFN-Laboratori Nazionali del Gran Sasso (LNGS), Via G. Acitelli 22, 67100 Assergi (AQ), Italy}

\corres{Correspondence: ; carmelo.evoli@gssi.it}

\abstract{We analyze the electron cosmic-ray spectrum from AMS-02, focusing on the spectral hardening around 42 GeV. Our findings confirm that this feature is intrinsic to the \emph{primary} electron component rather than a byproduct of contamination from primary positron sources. Even under conservative assumptions, its significance remains at about \(7\sigma\), strongly indicating a genuine spectral break. Accordingly, we introduce a new, more realistic parametric fit, which we recommend for the next round of AMS-02 analyses. Once the sources of systematic uncertainties are better constrained, this refined approach can either reinforce or refute our conclusions, providing a clearer understanding of the observed electron spectrum. If confirmed, we propose that this hardening most likely arises from interstellar transport or acceleration effects.}


\begin{document}

%%%%%%%%%%%%%%%%%%%%%%%%%%%%%%%%%%%%%%%%%%
\section{Introduction}
\label{sec:intro}

The AMS-02 (Alpha Magnetic Spectrometer) Collaboration has provided the most precise measurements of cosmic-ray electron and positron fluxes to date~\cite{electrons.AMS02,positrons.AMS02}. 
These data extend up to 1 TeV for positrons and 1.4 TeV for electrons, representing a significant advancement in our ability to study cosmic-ray particles at high energies. These measurements have confirmed that the fluxes of electrons and positrons deviate from predictions based purely on secondary production, prompting the development of various theoretical models to account for the observed discrepancies~\cite{DiMauro2014jcap,Yaun2015aph,Cholis2018prd,Evoli2021prd,Orusa2021jcap}.

One of the most intriguing findings from the AMS-02 data is the detection of a feature in the electron spectrum around \( 42.1^{+5.4}_{-5.2} \) GeV. This feature, characterized by a smooth hardening of the spectrum, was interpreted by the AMS-02 collaboration as evidence of at least two distinct electron components. According to their analysis, this feature can be modeled by overlapping two electron components: one with a very steep spectrum corresponding to a spectral index of \( \gamma_a = -4.31 \pm 0.13 \), and~another with a shallower slope of \( \gamma_b = -3.14 \pm 0.02 \)~\cite{electrons.AMS02}. 
The steep component, in~particular, challenges our current understanding of cosmic-ray sources. Because~these low-energy electrons experience relatively slow energy losses compared to higher-energy electrons, such a steep spectrum does not align with the characteristics of any known astrophysical origin~\cite{Li2015plb}.

Similar findings have been reported by the CALET (CALorimetric Electron Telescope)~\cite{leptons.CALET} and FERMI (Fermi Gamma-ray Space Telescope)~\cite{leptons.FERMI} collaborations, which also observed a hardening in the electron plus positron spectrum at comparable energies. However, the~energy uncertainties in these measurements are larger, and~they do not allow for the discrimination between positrons and electrons, leaving some ambiguity regarding the precise nature of the spectral~feature.

In our previous work~\cite{Evoli2020prl,Evoli2020prlerratum}, we proposed an alternative interpretation, suggesting that this spectral hardening could be linked to electron energy losses, specifically due to the onset of Klein-Nishina (KN) effects in the inverse Compton scattering (ICS) cross-section of electrons interacting with ultraviolet (UV) photons in the interstellar medium (ISM). The~KN effect, which becomes significant at high energies, reduces the efficiency of energy transfer during electron-photon interactions, potentially leading to the observed spectral hardening.
To arrive at this conclusion, however, we had to make somewhat ad-hoc assumptions regarding transport parameters, such as the average magnetic field in the diffusion zone and the interstellar radiation field (ISRF) in the inner Galaxy. Although~these choices fall within existing astrophysical uncertainties (see discussion in~{\cite{Evoli2021prd}), a~comprehensive parametric study is needed to conclusively confirm or rule out this scenario.
However, if~this interpretation is correct, the~detection of this feature would provide the most compelling evidence to date that electron transport in the Galaxy is dominated by energy losses rather than escape mechanisms. This has profound implications for our understanding of cosmic-ray propagation. Alternative models of cosmic-ray transport have been proposed~\cite{Blum2013prl,Cowsik2016apj,Lipari2017prd,Lipari2019prd}, in~which a very short escape time for leptons is invoked, with~energy losses becoming significant only at energies above several hundred GeV. However, the~presence of this feature at much lower energies challenges this view, suggesting that energy losses play a critical role in shaping the electron spectrum at energies around 40 GeV, consistent with the expectations of the standard halo model for the transport of Galactic cosmic rays~\cite{Strong2007arnps}.

At these energies, the~electron spectrum is believed to consist of contributions from primary electrons---likely accelerated alongside nuclei in localized astrophysical events such as Galactic supernova remnants (SNRs)---and a second component associated with the symmetric counterpart of the primary positron source~\cite{DiMauro2014jcap,Yaun2015aph,Cholis2018prd,Evoli2021prd,Orusa2021jcap}. The~hardening in the electron spectrum could, therefore, be induced by the presence of this second~component.

In this work, we present a novel analysis aimed at demonstrating that the feature observed by AMS-02 is not a result of contamination from electrons associated with the primary positron source population. Instead, we confirm that this spectral feature is intrinsic to the primary electron component itself, necessitating an explanation rooted in either the injection processes at supernova remnants or the subsequent transport mechanisms within the ISM, such as the influence of KN~effects.

%%%%%%%%%%%%%%%%%%%%%%%%%%%%%%%%%%%%%%%%%%
\section{Methods}
\label{sec:methods}

We utilized the precision measurements of the positron and electron absolute fluxes collected by AMS-02 during the period from 2011 to 2018, as~reported in~\cite{results.AMS02}.  
Both datasets were derived using the same analysis methodology and energy bins, which is crucial for the present~analysis.

For each energy bin, the~mean energy was computed following the method described in~\cite{Derome2019aa}, using the expression:
\begin{equation}\label{eq:emean}
\tilde{E}_i = \frac{\int_{E_{\rm min, i}}^{E_{\rm max, i}} dx \, x^{1-\gamma}}{\int_{E_{\rm min, i}}^{E_{\rm max, i}} dx \, x^{-\gamma}}~,
\end{equation}
where \( E_{\text{min},i} \) and \( E_{\text{max},i} \) are the minimum and maximum energies of the \( i \)-th bin, respectively, and~\( \gamma \) is the average slope of the flux within each bin. For~both datasets, we assumed a slope of \( \gamma = 3 \) and verified that varying the slope within the range of 2.5 to 3.5 does not appreciably impact the final~results.

Our analysis focuses on energies above 20 GeV, where the effects of solar modulation are negligible based on the analysis of~\cite{modulation.AMS02}, and~the contribution of secondary electrons is also minimal~\cite{Moskalenko1998apj}.

It is important to note that in the reference scenario, the~observed flux in this energy range is expected to be the result of contributions from multiple sources, and~thus we do not anticipate any irregular behavior due to individual local sources. Specifically, the~cooling timescale of electrons/positrons is given by \( \tau_{\rm cool} \sim 10 \, \text{Myr} \times (E / 40 \, \text{GeV})^{-1} \), while the diffusion length is \( \lambda_d \sim \sqrt{4 D \tau_{\rm cool}} \sim 4.5 \, \text{kpc} \times (E / 40 \, \text{GeV})^{-1/4} \). The~sources that contribute to the flux at Earth at a given energy \( E \) are those within a distance where the propagation time is shorter than the loss time at that energy. Assuming uniformly distributed sources in the Galactic disk with radius \( R_G \), the~number of sources within a distance \( \lambda_d \) from Earth, exploding within a loss timescale \( \tau_{\text{cool}} \), is approximately \( N \sim (\lambda_d / R_G)^2 \mathcal R \tau_{\rm cool} \). For~\( R_G \sim 15 \, \text{kpc} \) and  \( \mathcal R \sim 2 /\)century, this estimate yields \( \sim 2 \times 10^4 \) sources contributing at 40 GeV. 
Following this argument, a~scenario where \(\mathcal O(10)\) sources contribute to the local flux is plausible only at energies higher than 10 TeV, which exceeds the maximum energy currently measured by AMS-02. Dedicated simulations as those reported in~\cite{Evoli2021prd,Evoli2021prdb} confirm this scenario.

In the energy range considered here, the~positron flux is dominated by a primary component of positrons within the Galaxy, a~requirement to account for the increasing positron fraction first revealed by PAMELA~\cite{positrons.PAMELA}. This primary component is often associated with Galactic pulsars, which provide a successful explanation of the data~\cite{DiMauro2014jcap,Yaun2015aph,Cholis2018prd,Evoli2021prd,Orusa2021jcap}.

The emission from pulsars is assumed to be symmetric with respect to the production of electrons and positrons, thereby suggesting an equivalent flux of electrons contributing to the AMS-02 electron flux~\cite{Amato2018asr}.

In this scenario, the~local electron flux is the result of two contributions:
\begin{equation}
\Phi_{e^-}^{\rm obs} = \Phi_{e^-}^{\rm pri} + \Phi_{e^-}^{\rm X}~,
\end{equation}
where \( \Phi_{e^-}^X \) represents the counterpart of the primary positron source:  \(  \Phi_{e^-}^{\rm X}  =  \Phi_{e^+}^{\rm X} \)
Since our focus is on modeling the primary electron component, we make the reasonable assumption that the secondary contribution is symmetric with respect to the positron flux. As reported in literature, the~positrons typically favored by around 40\%~\cite{Orusa2022prd}. However, because~the secondary contribution to the electron flux is significantly subdominant, this imbalance has only a marginal impact on my overall analysis. 

Therefore, for~the purposes of this study, we concentrate on the data for \( e^- - e^+ \), which is dominated by the propagated primary cosmic-ray electrons. This approach minimizes potential uncertainties arising from the introduction of primary sources responsible for the positron excess, allowing us to remain agnostic about the actual origin of this~excess.

In Figure~\ref{fig:data}, we present the flux of electrons with positrons subtracted. The~error bars represent the statistical uncertainties, which are assumed to be the sum of the uncertainties for the two fluxes:
\begin{equation}\label{eq:sigmas}
\sigma_{e^- - e^+} = \sqrt{\sigma_{e^+}^2 + \sigma_{e^-}^2}~.
\end{equation}

In the next section, we will quantify the compatibility of this observable with a unique power-law spectrum. Any deviation from this behavior would imply that the observed feature arises from effects related to either the acceleration or the transport of the primary electron~component.

\begin{figure}[t]
%\centering
\includegraphics[width=.6\textwidth]{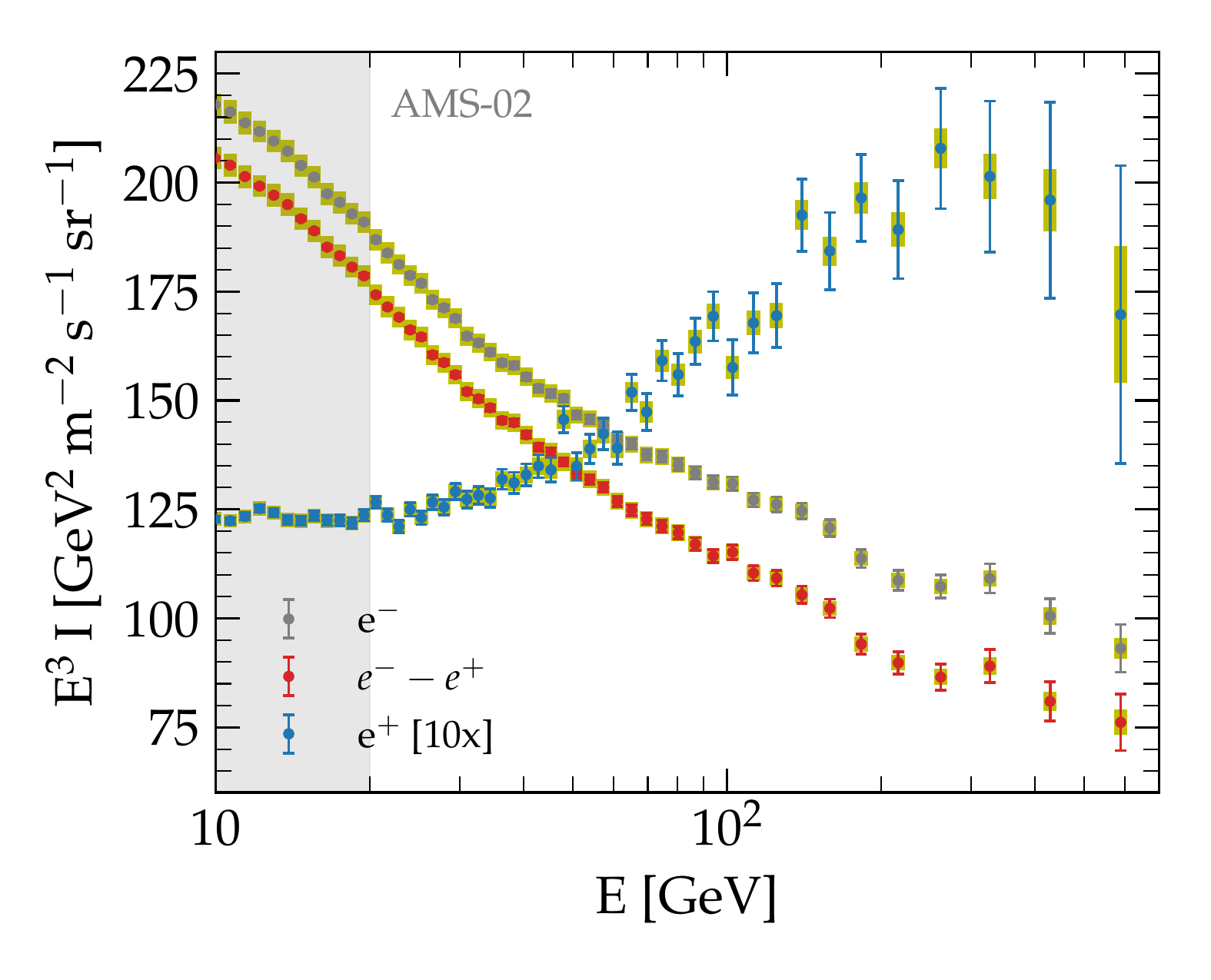}
\caption{The AMS-02 electron spectrum (\(\tilde{E}^3 \Phi_{e^-}\), gray data points) and positron spectrum (\(\tilde{E}^3 \Phi_{e^+}\), blue data points). The~red data points represent the flux of electrons minus the positron contribution (\(\tilde{E}^3 [\Phi_{e^-} - \Phi_{e^+}]\)). Statistical uncertainties are shown as error bars, while the yellow boxes represent systematic uncertainties. The shaded region indicates the energy range affected by solar~modulation.\label{fig:data}}
\end{figure}

%%%%%%%%%%%%%%%%%%%%%%%%%%%%%%%%%%%%%%%%%%
\section{Results}

\subsection{Electron-minus-positron spectrum fit}
\label{sec:empfit}

To quantify the spectral behavior of the \( e^- - e^+ \) spectrum, we employ two phenomenological models: a single power-law (SPL) function and a smoothly broken power-law (SBPL)~function.

The power-law model is expressed as:
\begin{equation}
F(E) = F_0 \left( \frac{E}{20~\text{GeV}} \right)^{-\alpha}~,
\end{equation}
where \( F_0 \) denotes the flux amplitude and \( \alpha \) represents the spectral index of the primary electrons.
The smoothly broken power-law model is given by:
\begin{equation}
F(E) = F_0 \left( \frac{E}{20~\text{GeV}} \right)^{-\alpha} \left[ 1 + \left( \frac{E}{E_b} \right)^s \right]^{\Delta \alpha / s}~.
\end{equation}
In this expression, \( E_b \) is the location of the spectral break, \( \Delta \alpha \) is the difference between the spectral indices before and after the break, \( s \) describes the smoothness of the break, and~\( F_0 \) is the flux amplitude as in the power-law~model.

Our goal is to compare these two models in their ability to describe the electron flux above 20 GeV. To~achieve this, we perform a \(\chi^2\) minimization using the AMS data on the cosmic-ray electron-minus-positron flux~measurements.

The \(\chi^2\) function is defined as:
\begin{equation}
\chi^2 = \sum_{i=1}^n \left( \frac{F_i - F(E_i)}{\sigma_{\text{stat},i}} \right)^2~,
\end{equation}
where, for~the \( i \)-th energy bin, \( E_i \), is the mean energy as defined in Equation~\eqref{eq:emean}, \( F_i \) is the electron-minus-positron measured flux, and~\( \sigma_{\text{stat},i} \) the statistical uncertainty as in \mbox{Equation~\eqref{eq:sigmas}},  and~\( F(E_i) \) is the flux predicted by the model for the corresponding energy~bin.

The best-fit parameters of the SPL and SBPL models are summarized in Table~\ref{tab:comparison}, along with the corresponding minimal \(\chi^2\) values and degrees of freedom. The \emph{p}-values are also listed, computed as the survival function (upper-tail probability) of the chi-square distribution with the specified degrees of freedom.

\begin{table}[H] 
\caption{Comparison of best-fit parameters for the two models described in Section~\ref{sec:empfit}.\label{tab:comparison}}
\newcolumntype{C}{>{\centering\arraybackslash}X}
\begin{tabularx}{\textwidth}{CCC}
\toprule
\textbf{Parameter}	     & \textbf{SPL Model}	 & \textbf{SBPL Model}\\
\midrule
$F_0$                    & $21.80 \pm 0.03$      & $22.02 \pm 0.05$ \\
$\alpha$                 & $3.280 \pm 0.002$     & $3.321 \pm 0.010$ \\
$E_b$                    & -                    & $37.2 \pm 3.6$ \\
$\Delta \alpha$          & -                   & $0.08 \pm 0.02$ \\
$s$                      & -                    & $0.007 \pm 0.009$ \\
$\chi^2/\mathrm{d.o.f.}$ & $96/35$               & $25.2/32$ \\
\midrule
\emph{p}-value                  & $1.31 \times 10^{-7}$ & $0.80$ \\
\bottomrule
\end{tabularx}
%\noindent{\footnotesize{\textsuperscript{1} Tables may have a footer.}}
\end{table}

The reduction of the \( \chi^2 \) value is therefore 71 for three more free parameters, suggesting a significance of $\sim 7.8~\sigma$ in favor of the SBPL model compared with the SPL~model.

The comparison of the best fit result for the SPL (dashed line) and SBPL (solid line) with the data is shown in Figure~\ref{fig:bestfit}.

\begin{figure}[t]
%\centering
\includegraphics[width=.6\textwidth]{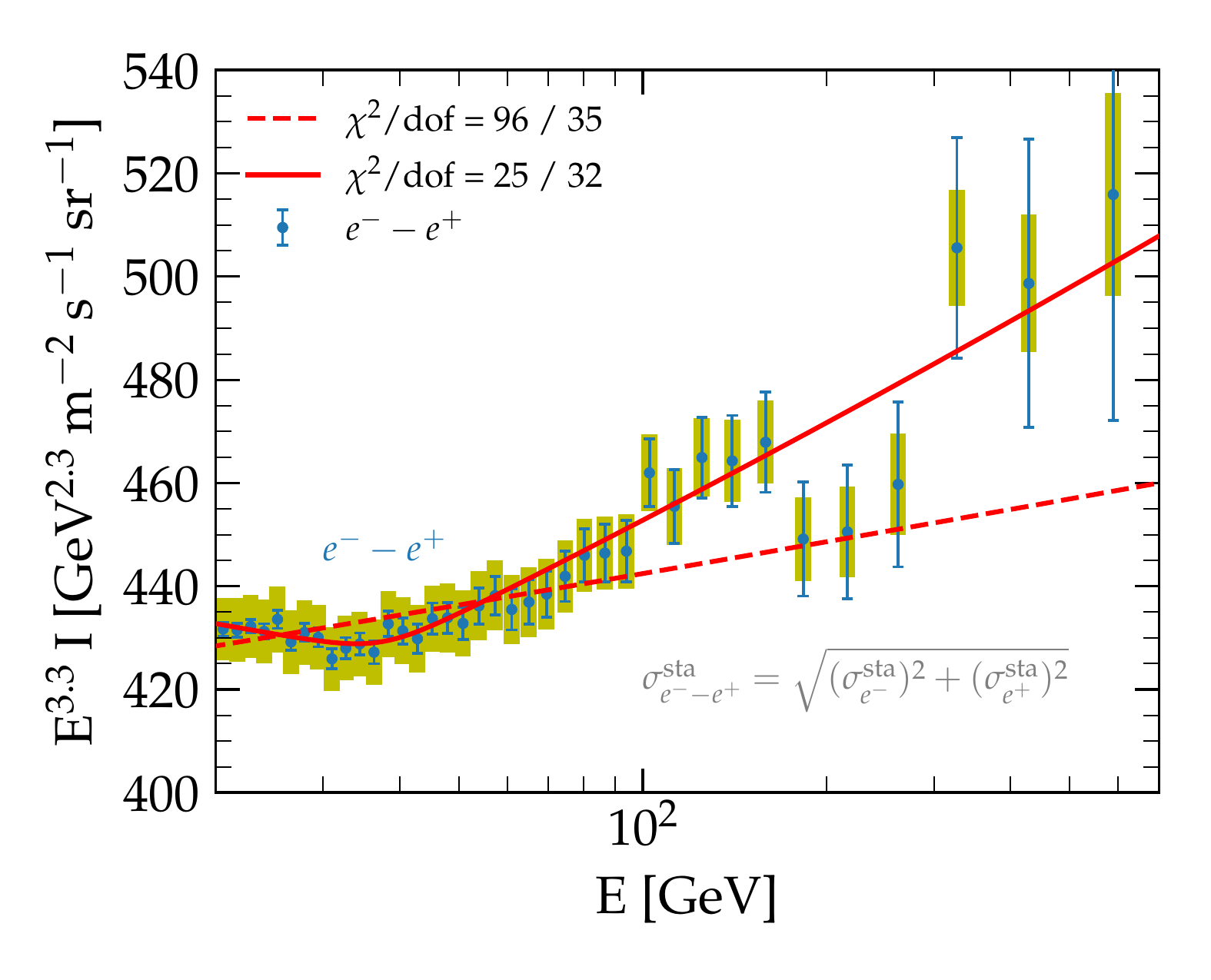}
\caption{Best-fit results for the electron-minus-positron spectrum, using the single power-law (SPL) model (dashed line) and the smoothly broken power-law (SBPL) model (solid line). The~corresponding $\chi^2$/d.o.f. values are indicated in the~legend.\label{fig:bestfit}}
\end{figure}

\subsection{Systematic uncertainties}
\label{sec:empfitsys}

The systematic uncertainties quoted by the AMS Collaboration vary as a function of energy, ranging between 3\% and 17\%. Most of these systematic uncertainties are correlated across energy bins, as~inferred from the description of the sources of these uncertainties in the corresponding publications. However, a~proper treatment of these correlations would require access to the correlation matrix, which has not been made publicly~available.

To assess the potential impact of these systematic uncertainties on our analysis, we considered an extreme scenario. Specifically, we examined the case where the maximum positron flux---given by its measured value plus the systematic uncertainty---is subtracted from the minimum electron flux, defined as the measured electron flux minus its systematic uncertainty. This approach allows us to test the robustness of our fitting results under the most conservative assumption regarding the contribution of the primary positron~source.

Despite this stringent test, we find that the fitting parameters for the PL and SBPL models change very little. Importantly, the~significance in favor of the SBPL model decreases slightly but remains strong at approximately 7 \(\sigma\) (see Table~\ref{tab:comparisonsys}). This result further strengthens the argument that the observed spectral hardening is not an artifact of systematic uncertainties, but~rather a genuine feature of the electron~spectrum.

\begin{table}[H] 
\caption{Comparison of best-fit parameters for the two models described in Section~\ref{sec:empfitsys}.\label{tab:comparisonsys}}
\newcolumntype{C}{>{\centering\arraybackslash}X}
\begin{tabularx}{\textwidth}{CCCCC}
\toprule
\textbf{Parameter} & \textbf{SPL Model}	& \textbf{SBPL Model} & \textbf{SPL Model} & \textbf{SBPL Model} \\
& \multicolumn{2}{c}{\textbf{Correlated Uncertainties}} & \multicolumn{2}{c}{\textbf{Uncorrelated Uncertainties}} \\
\midrule
$F_0$                    & $21.48 \pm 0.03$  & $21.69 \pm 0.05$  & $21.28 \pm 0.11$  & $21.70 \pm 0.22$ \\
$\alpha$                 & $3.282 \pm 0.002$ & $3.323 \pm 0.009$ & $3.268 \pm 0.005$ & $3.326 \pm 0.046$ \\
$E_b$                    & -                & $36.3 \pm  3.7$   & -                & $36.1 \pm 9.6$ \\
$\Delta \alpha$          & -                & $0.07 \pm 0.012$  & -                & $0.08 \pm 0.05$ \\
$s$                      & -                & $0.005 \pm 0.009$ & -                & $0.007 \pm 0.020$ \\
$\chi^2/\mathrm{d.o.f.}$ & $89/35$           & $24.2/32$         & $18/35$           & $8.6/32$ \\
\midrule
\emph{p}-value                  & $1.35 \times 10^{-6}$ & $0.84$        & $0.99$ & $1.00$  \\
\bottomrule
\end{tabularx}
%\noindent{\footnotesize{\textsuperscript{1} Tables may have a footer.}}
\end{table}

We also examined the opposite approach, treating all systematic uncertainties as uncorrelated by adding them in quadrature to the statistical ones. Although~this method is sometimes used when details of the uncertainties are lacking, it is widely recognized as overly simplistic and can likely overestimates the impact of systematics. We find that the best-fit parameters remain practically unchanged, indicating that the energy dependence of the systematic uncertainties is weak and does not significantly affect our analysis. As~expected, the~main effect is a reduction in the best-fit \(\chi^2\). Consequently, the~significance of the spectral feature decreases but remains at the \(2\sigma\) level~(see Table~\ref{tab:comparisonsys}).

\subsection{Positron flux modelling}
\label{sec:empfitnew}

An alternative method we explored is to adopt a minimal phenomenological model for the positron flux. The~AMS Collaboration introduced the following functional form,
\begin{equation}
B(E) = C_1 \left( \frac{E}{20~\text{GeV}} \right)^{-\gamma_1} 
      + C_2 \left( \frac{E}{E_2} \right)^{-\gamma_2} \exp \left( - \frac{E}{E_C} \right)~,
\end{equation}
which was shown to describe their data with high accuracy over the full energy range. This model includes the minimal number of parameters required to fit the positron~data.

Here, \( \gamma_i \) and \( C_i \) (\( i \in \{1,2\} \)) denote the spectral indices and normalizations for two distinct positron populations, \( E_2 \) marks the energy at which the second population becomes significant, and~\( E_C \) is the cutoff energy. We determine the initial parameters of this model through a \(\chi^2\)-minimization using the absolute positron flux data from AMS. The~resulting best-fit parameters are \( C_1 = 0.6 \pm 0.1 \), \( \gamma_1 = 3.9 \pm 0.2 \), \( E_b = 53.8  \pm 1.1 \)~GeV, \( C_2 = 0.085 \pm 0.004 \), \( \gamma_2 = 2.50 \pm 0.05 \), \( E_c =  641 \pm 126 \)~GeV, yielding a minimal \(\chi^2 / \text{d.o.f.} = 38 / 31\), corresponding to a \(p\)-value of 0.182. The~resulting fit is shown in Figure~\ref{fig:databackground}.

Having established this description of the positron flux, we then add it to either the SPL or BPL model for electrons. In~practice, the~\(\chi^2\) function to be minimized is
\begin{equation}
\chi^2 = \sum_{i \in e^+} \left( \frac{F_i - B(E_i)}{\sigma_{\text{stat},i}} \right)^2 \;+\;  
         \sum_{i \in e^-} \left( \frac{F_i - \bigl[F(E_i) + B(E_i)\bigr]}{\sigma_{\text{stat},i}} \right)^2~,
\end{equation}
where \(B(E)\) models the positrons, and~\(F(E)\) is either of the two electron models introduced previously (SPL or BPL). We summarize the corresponding best-fit parameters in Table~\ref{tab:comparisonnew} and display the two models in Figure~\ref{fig:databackground}.

In this case, the~\(\chi^2\) decreases by 70 upon adding three free parameters, indicating a significance of approximately \(7.7\sigma\) in favor of the SBPL scenario relative to the PL model. Finally, we note that if we add systematic uncertainties in quadrature with the statistical ones, effectively treating them as uncorrelated, the~significance of the feature is reduced to \(2.2\sigma\). As~discussed in the previous section, this underlines the critical importance of accurately characterizing systematic uncertainties and their correlations before drawing a definitive conclusion on the presence of the feature.

\begin{table}[H] 
\caption{Comparison of best-fit parameters for the two models described in Section~\ref{sec:empfitnew}.\label{tab:comparisonnew}}
\newcolumntype{C}{>{\centering\arraybackslash}X}
\begin{tabularx}{\textwidth}{CCC}
\toprule
\textbf{Parameter}	& \textbf{SPL Model}	& \textbf{SBPL Model}\\
\midrule
C1                    & $0.6 \pm 0.1$       & $0.6 \pm 0.1$         \\
$\alpha_1$            & $4.1 \pm 0.2$       & $3.8 \pm 0.1$         \\
$E_{b1}$              & $53.8 \pm 1.0$      & $53.9 \pm 1.0$        \\
C2                    & $0.086 \pm 0.004$   & $0.084 \pm 0.004$     \\
$\alpha_2$            & $2.48 \pm 0.04$     & $2.49 \pm 0.04$       \\
$E_c$                 & $668 \pm 120$       & $641 \pm 115$         \\
$I_0$                 & $21.80 \pm 0.03$    & $22.02 \pm 0.04$      \\
$\alpha$              & $3.28 \pm 0.01$     & $3.32 \pm 0.01$       \\
$E_b$                 & -                  & $37.58 \pm 3.43$      \\
$\Delta \alpha$       & -                  & $0.08 \pm 0.02$       \\
$s$                   & -                  & $0.01 \pm 0.01$       \\
$\chi^2/\mathrm{dof}$ & $136/67$            & $66/64$               \\
\midrule
\emph{p}-value               & $1.369\times 10^{-6}$ & $0.394$             \\
\bottomrule
\end{tabularx}
%\noindent{\footnotesize{\textsuperscript{1} Tables may have a footer.}}
\end{table}
\vspace{-6pt}
\begin{figure}[H]
%\centering
\includegraphics[width=.495\textwidth]{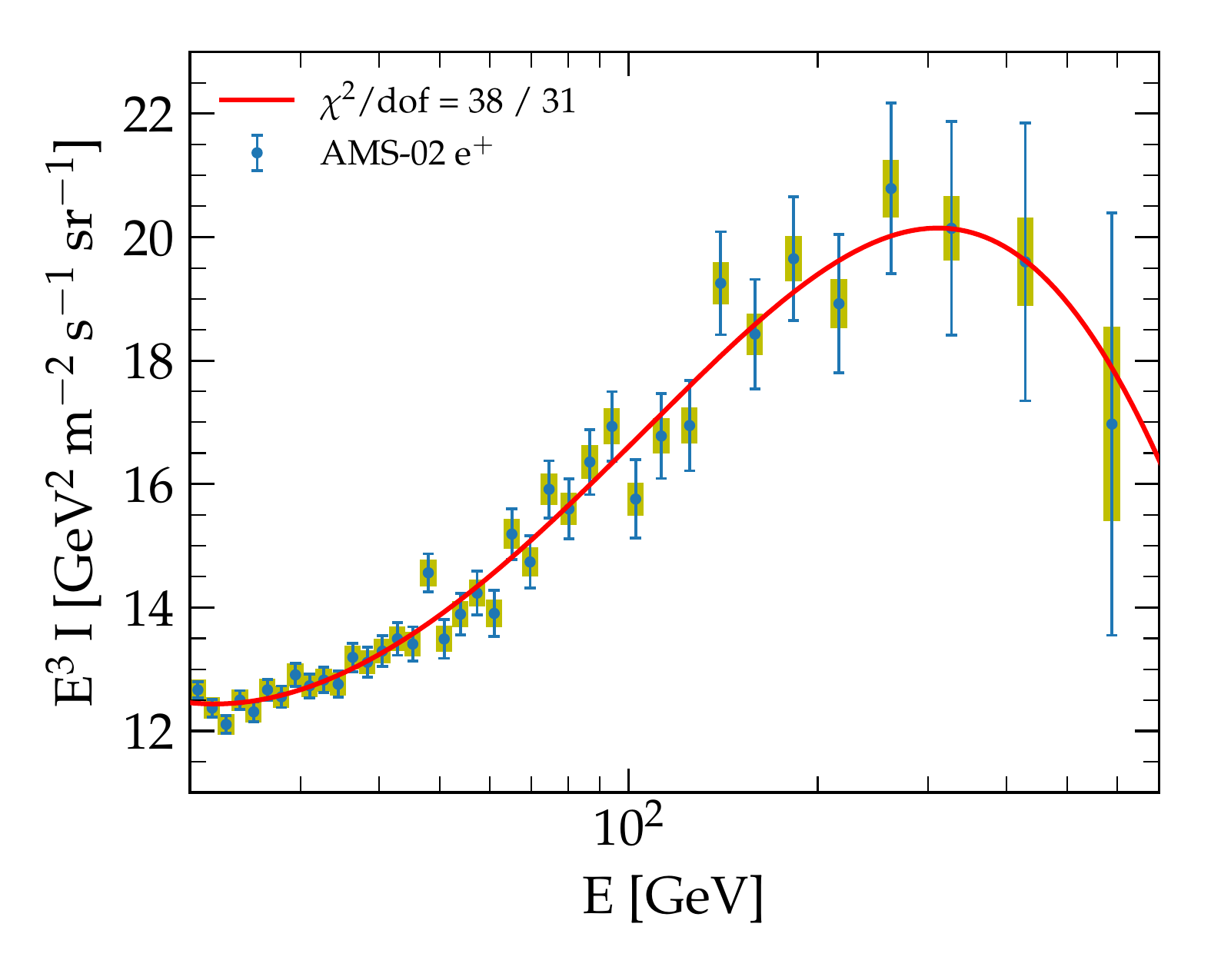}
\includegraphics[width=.495\textwidth]{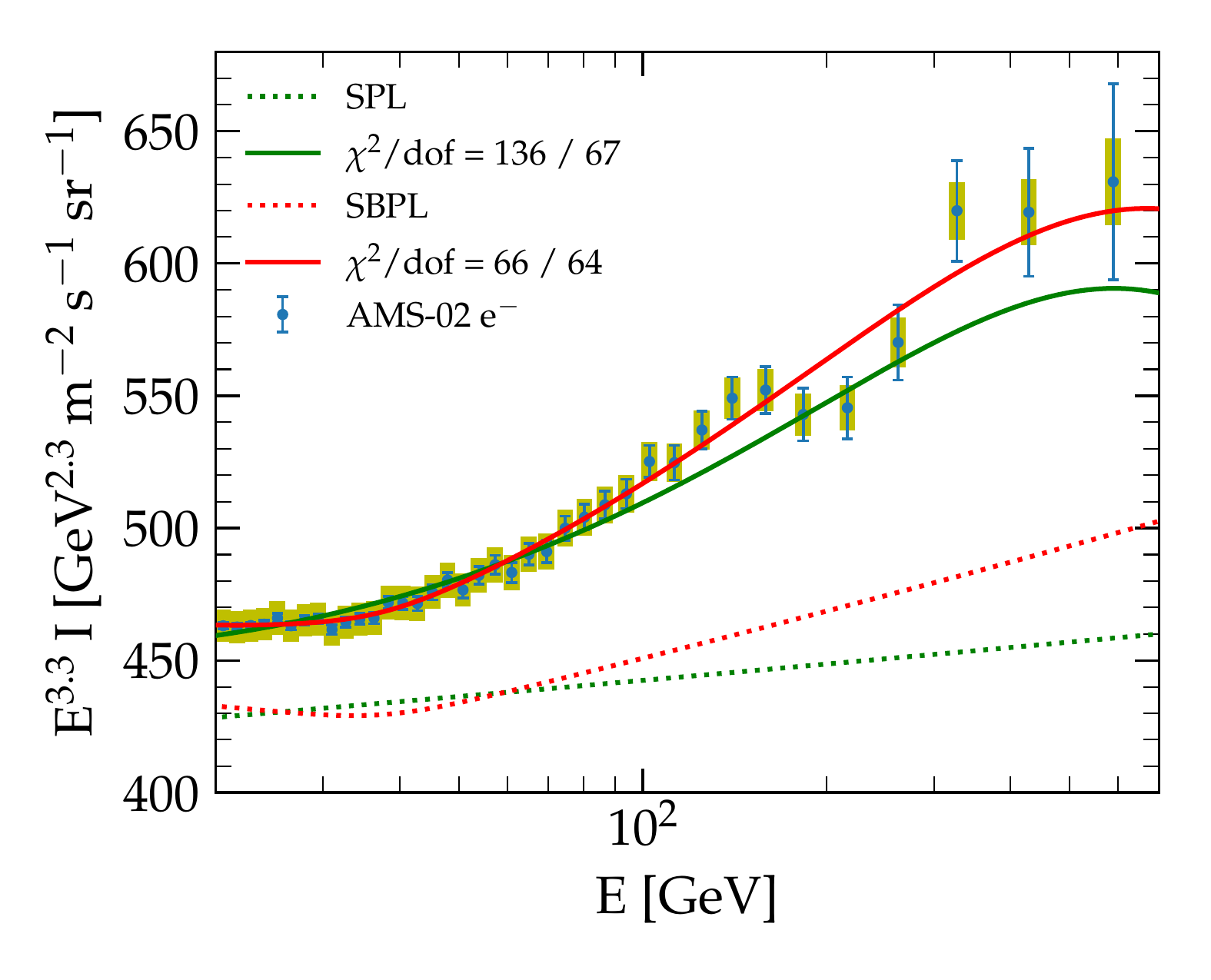}
\caption{(\textbf{Left}) The AMS-02 positron spectrum (blue data points) compared with the best-fit phenomenological model \(B\) (dashed red line) introduced in Section~\ref{sec:empfit}. (\textbf{Right}) The AMS-02 electron data (blue data points) compared with the best-fit BPL\(+B\) (dashed green line) and SBPL\(+B\) (dashed red line). The~dotted lines show the corresponding fits without the additional positron (symmetric) component \(B\). \label{fig:databackground}}
\end{figure}

\section{Conclusions}

In this study, we have analyzed the electron cosmic-ray spectrum using data from the AMS-02 experiment, focusing on understanding the spectral feature observed around \( 42.1^{+5.4}_{-5.2} \) GeV. Our results confirm that this feature is not due to contamination from electrons associated with the primary positron source population but is intrinsic to the primary electron component~itself.

This analysis was made possible by the release of the AMS-02 electron/positron spectra with unprecedented accuracy over a wide energy range~\cite{electrons.AMS02,positrons.AMS02}.

We assessed the robustness of our findings by considering the impact of systematic uncertainties in the AMS-02 data. Even under the most conservative assumptions—where the maximum positron flux plus its systematic uncertainty was subtracted from the minimum electron flux—the significance of the spectral hardening remained strong at approximately 7 \(\sigma\). This result reinforces the conclusion that the spectral break is a genuine feature of the \emph{primary} electron~spectrum.

A critical assumption in our analysis is that the sources contributing to the positron excess do not generate a significantly higher abundance of electrons at corresponding energies. This assumption holds for the most widely discussed cosmic-ray electron/positron sources, such as pulsars and dark matter annihilation or decay~\cite{Cavasonza2017apj}.

This feature cannot be attributed to solar modulation for several reasons. First, the~variance in the electron flux above 20 GeV measured by AMS-02 over nearly one solar cycle is smaller than the excess residual~\cite{modulation.AMS02}. More importantly, solar modulation typically produces a low-energy cutoff, which would have the opposite effect, further enhancing the significance of the observed excess if present~\cite{Cholis2022prd}.

Therefore, the~persistence of this feature supports the hypothesis that it originates from either injection processes at supernova remnants or subsequent transport~mechanisms. 

In the first scenario, the~spectral hardening could result from the injection spectrum hardening at high energies, as~expected in the non-linear cosmic-ray acceleration model~\cite{Ptuskin2013apj,Recchia2018mnras}, or~from the superposition of variable injection spectra from different cosmic-ray sources~\cite{Yuan2011prd}, where some sources may accelerate cosmic rays with harder spectra than usual.
In the second scenario, second-order Fermi acceleration in the interstellar medium could play a role in hardening the low-energy part of the cosmic-ray~spectrum. 

However, in~both scenarios, similar excesses should appear in the spectra of cosmic-ray nuclei, which are not observed at this energy~scale.

In~\cite{Evoli2020prl}, a~satisfactory description of the electron spectrum is achieved by considering leptonic energy losses that include the Klein-Nishina transition on the UV galactic photon background.
Remarkably, this explanation do not imply a similar spectral hardening in~nuclei.

If confirmed, our findings challenge alternative cosmic-ray transport models that rely on very short escape times for leptons and suggest that energy losses play a critical role in shaping the electron spectrum at energies around 40~GeV.

These conclusions contribute to a deeper understanding of cosmic-ray propagation in our Galaxy and underscore the need for experiments like AMS-02 to conduct more detailed analyses of measurements in this energy range, to~more robustly assess the properties of spectral features in the cosmic-ray~spectrum.

\funding{This work has been partially funded by the European Union – Next Generation EU, through PRIN-MUR 2022TJW4EJ and by the European Union – NextGenerationEU under the MUR National Innovation Ecosystem grant ECS00000041 – VITALITY/ASTRA – CUP D13C21000430001.}

\acknowledgments{I thank Luciana Andrade Dourado and Giovanni Benato for useful discussions. I gratefully acknowledge the use of the Cosmic-Ray DataBase (\href{https://lpsc.in2p3.fr/crdb/}{CRDB}), for accessing cosmic-ray data, and we appreciate the efforts of the CRDB team in providing this comprehensive and user-friendly database~\cite{Maurin2023crdb}. During the preparation of this manuscript, the author used ChatGPT-o1 for the purposes of improving syntax and grammar checking. The author has reviewed and edited the output and take full responsibility for the content of this publication.}

%\dataavailability{The original code presented in the study is openly available at \url{https://github.com/carmeloevoli/CosmicRayLeptonBreaks/tree/main/electrons_ams02}.} 

\reftitle{References}

\bibliography{biblio}

\begin{thebibliography}{999}

\bibitem[{Aguilar} et~al.(2019{\natexlab{a}}){Aguilar}, {Ali Cavasonza}, {Alpat}, {Ambrosi}, {Arruda}, {Attig}, {Azzarello}, {Bachlechner}, {Barao}, {Barrau}, and et~al.]{electrons.AMS02}
{Aguilar}, M.; {Ali Cavasonza}, L.; {Alpat}, B.; {Ambrosi}, G.; {Arruda}, L.; {Attig}, N.; {Azzarello}, P.; {Bachlechner}, A.; {Barao}, F.; {Barrau}, A.;  et~al.
\newblock {Towards Understanding the Origin of Cosmic-Ray Electrons}.
\newblock {\em \prl} {\bf 2019}, {\em 122},~101101.
\newblock {\url{https://doi.org/10.1103/PhysRevLett.122.101101}}.

\bibitem[{Aguilar} et~al.(2019{\natexlab{b}}){Aguilar}, {Ali Cavasonza}, {Ambrosi}, {Arruda}, {Attig}, {Azzarello}, {Bachlechner}, {Barao}, {Barrau}, {Barrin}, and et~al.]{positrons.AMS02}
{Aguilar}, M.; {Ali Cavasonza}, L.; {Ambrosi}, G.; {Arruda}, L.; {Attig}, N.; {Azzarello}, P.; {Bachlechner}, A.; {Barao}, F.; {Barrau}, A.; {Barrin}, L.;  et~al.
\newblock {Towards Understanding the Origin of Cosmic-Ray Positrons}.
\newblock {\em \prl} {\bf 2019}, {\em 122},~041102.
\newblock {\url{https://doi.org/10.1103/PhysRevLett.122.041102}}.

\bibitem[{Di Mauro} et~al.(2014){Di Mauro}, {Donato}, {Fornengo}, {Lineros}, and {Vittino}]{DiMauro2014jcap}
{Di Mauro}, M.; {Donato}, F.; {Fornengo}, N.; {Lineros}, R.; {Vittino}, A.
\newblock {Interpretation of AMS-02 electrons and positrons data}.
\newblock {\em \jcap} {\bf 2014}, {\em 2014},~006,  \href{http://arxiv.org/abs/1402.0321}{{\normalfont [arXiv:astro-ph.HE/1402.0321]}}.
\newblock {\url{https://doi.org/10.1088/1475-7516/2014/04/006}}.

\bibitem[{Yuan} et~al.(2015){Yuan}, {Bi}, {Chen}, {Guo}, {Lin}, and {Zhang}]{Yaun2015aph}
{Yuan}, Q.; {Bi}, X.J.; {Chen}, G.M.; {Guo}, Y.Q.; {Lin}, S.J.; {Zhang}, X.
\newblock {Implications of the AMS-02 positron fraction in cosmic rays}.
\newblock {\em Astroparticle Physics} {\bf 2015}, {\em 60},~1--12,  \href{http://arxiv.org/abs/1304.1482}{{\normalfont [arXiv:astro-ph.HE/1304.1482]}}.
\newblock {\url{https://doi.org/10.1016/j.astropartphys.2014.05.005}}.

\bibitem[{Cholis} et~al.(2018){Cholis}, {Karwal}, and {Kamionkowski}]{Cholis2018prd}
{Cholis}, I.; {Karwal}, T.; {Kamionkowski}, M.
\newblock {Studying the Milky Way pulsar population with cosmic-ray leptons}.
\newblock {\em \prd} {\bf 2018}, {\em 98},~063008,  \href{http://arxiv.org/abs/1807.05230}{{\normalfont [arXiv:astro-ph.HE/1807.05230]}}.
\newblock {\url{https://doi.org/10.1103/PhysRevD.98.063008}}.

\bibitem[{Evoli} et~al.(2021){Evoli}, {Amato}, {Blasi}, and {Aloisio}]{Evoli2021prd}
{Evoli}, C.; {Amato}, E.; {Blasi}, P.; {Aloisio}, R.
\newblock {Galactic factories of cosmic-ray electrons and positrons}.
\newblock {\em \prd} {\bf 2021}, {\em 103},~083010,  \href{http://arxiv.org/abs/2010.11955}{{\normalfont [arXiv:astro-ph.HE/2010.11955]}}.
\newblock {\url{https://doi.org/10.1103/PhysRevD.103.083010}}.

\bibitem[{Orusa} et~al.(2021){Orusa}, {Manconi}, {Donato}, and {Di Mauro}]{Orusa2021jcap}
{Orusa}, L.; {Manconi}, S.; {Donato}, F.; {Di Mauro}, M.
\newblock {Constraining positron emission from pulsar populations with AMS-02 data}.
\newblock {\em \jcap} {\bf 2021}, {\em 2021},~014,  \href{http://arxiv.org/abs/2107.06300}{{\normalfont [arXiv:astro-ph.HE/2107.06300]}}.
\newblock {\url{https://doi.org/10.1088/1475-7516/2021/12/014}}.

\bibitem[{Li} et~al.(2015){Li}, {Shen}, {Lu}, {Dong}, {Fan}, {Feng}, {Liu}, and {Chang}]{Li2015plb}
{Li}, X.; {Shen}, Z.Q.; {Lu}, B.Q.; {Dong}, T.K.; {Fan}, Y.Z.; {Feng}, L.; {Liu}, S.M.; {Chang}, J.
\newblock {'Excess' of primary cosmic ray electrons}.
\newblock {\em Physics Letters B} {\bf 2015}, {\em 749},~267--271,  \href{http://arxiv.org/abs/1412.1550}{{\normalfont [arXiv:astro-ph.HE/1412.1550]}}.
\newblock {\url{https://doi.org/10.1016/j.physletb.2015.08.001}}.

\bibitem[{Adriani} et~al.(2023){Adriani}, {Akaike}, {Asano}, {Asaoka}, {Berti}, {Bigongiari}, {Binns}, {Bongi}, {Brogi}, {Bruno}, {Buckley}, {Cannady}, {Castellini}, {Checchia}, {Cherry}, {Collazuol}, {de Nolfo}, {Ebisawa}, {Ficklin}, {Fuke}, {Gonzi}, {Guzik}, {Hams}, {Hibino}, {Ichimura}, {Ioka}, {Ishizaki}, {Israel}, {Kasahara}, {Kataoka}, {Kataoka}, {Katayose}, {Kato}, {Kawanaka}, {Kawakubo}, {Kobayashi}, {Kohri}, {Krawczynski}, {Krizmanic}, {Maestro}, {Marrocchesi}, {Messineo}, {Mitchell}, {Miyake}, {Moiseev}, {Mori}, {Mori}, {Motz}, {Munakata}, {Nakahira}, {Nishimura}, {Okuno}, {Ormes}, {Ozawa}, {Pacini}, {Papini}, {Rauch}, {Ricciarini}, {Sakai}, {Sakamoto}, {Sasaki}, {Shimizu}, {Shiomi}, {Spillantini}, {Stolzi}, {Sugita}, {Sulaj}, {Takita}, {Tamura}, {Terasawa}, {Torii}, {Tsunesada}, {Uchihori}, {Vannuccini}, {Wefel}, {Yamaoka}, {Yanagita}, {Yoshida}, {Yoshida}, {Zober}, and {Calet Collaboration}]{leptons.CALET}
{Adriani}, O.; {Akaike}, Y.; {Asano}, K.; {Asaoka}, Y.; {Berti}, E.; {Bigongiari}, G.; {Binns}, W.R.; {Bongi}, M.; {Brogi}, P.; {Bruno}, A.;  et~al.
\newblock {Direct Measurement of the Spectral Structure of Cosmic-Ray Electrons+Positrons in the TeV Region with CALET on the International Space Station}.
\newblock {\em \prl} {\bf 2023}, {\em 131},~191001,  \href{http://arxiv.org/abs/2311.05916}{{\normalfont [arXiv:astro-ph.HE/2311.05916]}}.
\newblock {\url{https://doi.org/10.1103/PhysRevLett.131.191001}}.

\bibitem[{Abdollahi} et~al.(2017){Abdollahi}, {Ackermann}, {Ajello}, {Atwood}, {Baldini}, {Barbiellini}, {Bastieri}, {Bellazzini}, {Bloom}, {Bonino}, {Brandt}, {Bregeon}, {Bruel}, {Buehler}, {Cameron}, {Caputo}, {Caragiulo}, {Castro}, {Cavazzuti}, {Cecchi}, {Chekhtman}, {Ciprini}, {Cohen-Tanugi}, {Costanza}, {Cuoco}, {Cutini}, {D'Ammando}, {de Palma}, {Desiante}, {Digel}, {Di Lalla}, {Di Mauro}, {Di Venere}, {Drell}, {Drlica-Wagner}, {Favuzzi}, {Focke}, {Funk}, {Fusco}, {Gargano}, {Gasparrini}, {Giglietto}, {Giordano}, {Giroletti}, {Green}, {Guillemot}, {Guiriec}, {Harding}, {Jogler}, {J{\'o}hannesson}, {Kamae}, {Kuss}, {La Mura}, {Latronico}, {Longo}, {Loparco}, {Lubrano}, {Maldera}, {Malyshev}, {Manfreda}, {Mazziotta}, {Michelson}, {Mirabal}, {Mitthumsiri}, {Mizuno}, {Moiseev}, {Monzani}, {Morselli}, {Moskalenko}, {Negro}, {Nuss}, {Orlando}, {Paneque}, {Perkins}, {Pesce-Rollins}, {Piron}, {Pivato}, {Porter}, {Principe}, {Rain{\`o}}, {Rando}, {Razzano}, {Reimer}, {Reimer}, {Sgr{\`o}}, {Simone}, {Siskind},
  {Spada}, {Spandre}, {Spinelli}, {Tajima}, {Thayer}, {Tibaldo}, {Torres}, {Troja}, {Wood}, {Worley}, {Zaharijas}, {Zimmer}, and {Fermi-LAT Collaboration}]{leptons.FERMI}
{Abdollahi}, S.; {Ackermann}, M.; {Ajello}, M.; {Atwood}, W.B.; {Baldini}, L.; {Barbiellini}, G.; {Bastieri}, D.; {Bellazzini}, R.; {Bloom}, E.D.; {Bonino}, R.;  et~al.
\newblock {Cosmic-ray electron-positron spectrum from 7 GeV to 2 TeV with the Fermi Large Area Telescope}.
\newblock {\em \prd} {\bf 2017}, {\em 95},~082007,  \href{http://arxiv.org/abs/1704.07195}{{\normalfont [arXiv:astro-ph.HE/1704.07195]}}.
\newblock {\url{https://doi.org/10.1103/PhysRevD.95.082007}}.

\bibitem[{Evoli} et~al.(2020){Evoli}, {Blasi}, {Amato}, and {Aloisio}]{Evoli2020prl}
{Evoli}, C.; {Blasi}, P.; {Amato}, E.; {Aloisio}, R.
\newblock {Signature of Energy Losses on the Cosmic Ray Electron Spectrum}.
\newblock {\em \prl} {\bf 2020}, {\em 125},~051101,  \href{http://arxiv.org/abs/2007.01302}{{\normalfont [arXiv:astro-ph.HE/2007.01302]}}.
\newblock {\url{https://doi.org/10.1103/PhysRevLett.125.051101}}.

\bibitem[{Evoli} et~al.(2021){Evoli}, {Blasi}, {Amato}, and {Aloisio}]{Evoli2020prlerratum}
{Evoli}, C.; {Blasi}, P.; {Amato}, E.; {Aloisio}, R.
\newblock {Erratum: Signature of Energy Losses on the Cosmic Ray Electron Spectrum [Phys. Rev. Lett. 125, 051101 (2020)]}.
\newblock {\em \prl} {\bf 2021}, {\em 126},~249901.
\newblock {\url{https://doi.org/10.1103/PhysRevLett.126.249901}}.

\bibitem[{Blum} et~al.(2013){Blum}, {Katz}, and {Waxman}]{Blum2013prl}
{Blum}, K.; {Katz}, B.; {Waxman}, E.
\newblock {AMS-02 Results Support the Secondary Origin of Cosmic Ray Positrons}.
\newblock {\em \prl} {\bf 2013}, {\em 111},~211101,  \href{http://arxiv.org/abs/1305.1324}{{\normalfont [arXiv:astro-ph.HE/1305.1324]}}.
\newblock {\url{https://doi.org/10.1103/PhysRevLett.111.211101}}.

\bibitem[{Cowsik} and {Madziwa-Nussinov}(2016)]{Cowsik2016apj}
{Cowsik}, R.; {Madziwa-Nussinov}, T.
\newblock {Spectral Intensities of Antiprotons and the Nested Leaky-box Model for Cosmic Rays in the Galaxy}.
\newblock {\em \apj} {\bf 2016}, {\em 827},~119,  \href{http://arxiv.org/abs/1505.00305}{{\normalfont [arXiv:astro-ph.HE/1505.00305]}}.
\newblock {\url{https://doi.org/10.3847/0004-637X/827/2/119}}.

\bibitem[{Lipari}(2017)]{Lipari2017prd}
{Lipari}, P.
\newblock {Interpretation of the cosmic ray positron and antiproton fluxes}.
\newblock {\em \prd} {\bf 2017}, {\em 95},~063009,  \href{http://arxiv.org/abs/1608.02018}{{\normalfont [arXiv:astro-ph.HE/1608.02018]}}.
\newblock {\url{https://doi.org/10.1103/PhysRevD.95.063009}}.

\bibitem[{Lipari}(2019)]{Lipari2019prd}
{Lipari}, P.
\newblock {Spectral shapes of the fluxes of electrons and positrons and the average residence time of cosmic rays in the Galaxy}.
\newblock {\em \prd} {\bf 2019}, {\em 99},~043005,  \href{http://arxiv.org/abs/1810.03195}{{\normalfont [arXiv:astro-ph.HE/1810.03195]}}.
\newblock {\url{https://doi.org/10.1103/PhysRevD.99.043005}}.

\bibitem[{Strong} et~al.(2007){Strong}, {Moskalenko}, and {Ptuskin}]{Strong2007arnps}
{Strong}, A.W.; {Moskalenko}, I.V.; {Ptuskin}, V.S.
\newblock {Cosmic-Ray Propagation and Interactions in the Galaxy}.
\newblock {\em Annual Review of Nuclear and Particle Science} {\bf 2007}, {\em 57},~285--327,  \href{http://arxiv.org/abs/astro-ph/0701517}{{\normalfont [arXiv:astro-ph/astro-ph/0701517]}}.
\newblock {\url{https://doi.org/10.1146/annurev.nucl.57.090506.123011}}.

\bibitem[{Aguilar} et~al.(2021){Aguilar}, {Ali Cavasonza}, {Ambrosi}, {Arruda}, {Attig}, {Barao}, {Barrin}, {Bartoloni}, {Ba{\c{s}}e{\u{g}}mez-du Pree}, {Bates}, and et~al.]{results.AMS02}
{Aguilar}, M.; {Ali Cavasonza}, L.; {Ambrosi}, G.; {Arruda}, L.; {Attig}, N.; {Barao}, F.; {Barrin}, L.; {Bartoloni}, A.; {Ba{\c{s}}e{\u{g}}mez-du Pree}, S.; {Bates}, J.;  et~al.
\newblock {The Alpha Magnetic Spectrometer (AMS) on the international space station: Part II - Results from the first seven years}.
\newblock {\em \physrep} {\bf 2021}, {\em 894},~1--116.
\newblock {\url{https://doi.org/10.1016/j.physrep.2020.09.003}}.

\bibitem[{Derome} et~al.(2019){Derome}, {Maurin}, {Salati}, {Boudaud}, {G{\'e}nolini}, and {Kunz{\'e}}]{Derome2019aa}
{Derome}, L.; {Maurin}, D.; {Salati}, P.; {Boudaud}, M.; {G{\'e}nolini}, Y.; {Kunz{\'e}}, P.
\newblock {Fitting B/C cosmic-ray data in the AMS-02 era: a cookbook. Model numerical precision, data covariance matrix of errors, cross-section nuisance parameters, and mock data}.
\newblock {\em \aap} {\bf 2019}, {\em 627},~A158,  \href{http://arxiv.org/abs/1904.08210}{{\normalfont [arXiv:astro-ph.HE/1904.08210]}}.
\newblock {\url{https://doi.org/10.1051/0004-6361/201935717}}.

\bibitem[{Aguilar} et~al.(2018){Aguilar}, {Cavasonza}, {Ambrosi}, {Arruda}, {Attig}, {Aupetit}, {Azzarello}, {Bachlechner}, {Barao}, {Barrau}, and et~al.]{modulation.AMS02}
{Aguilar}, M.; {Cavasonza}, L.A.; {Ambrosi}, G.; {Arruda}, L.; {Attig}, N.; {Aupetit}, S.; {Azzarello}, P.; {Bachlechner}, A.; {Barao}, F.; {Barrau}, A.;  et~al.
\newblock {Observation of Complex Time Structures in the Cosmic-Ray Electron and Positron Fluxes with the Alpha Magnetic Spectrometer on the International Space Station}.
\newblock {\em \prl} {\bf 2018}, {\em 121},~051102.
\newblock {\url{https://doi.org/10.1103/PhysRevLett.121.051102}}.

\bibitem[{Moskalenko} and {Strong}(1998)]{Moskalenko1998apj}
{Moskalenko}, I.V.; {Strong}, A.W.
\newblock {Production and Propagation of Cosmic-Ray Positrons and Electrons}.
\newblock {\em \apj} {\bf 1998}, {\em 493},~694--707,  \href{http://arxiv.org/abs/astro-ph/9710124}{{\normalfont [arXiv:astro-ph/astro-ph/9710124]}}.
\newblock {\url{https://doi.org/10.1086/305152}}.

\bibitem[{Evoli} et~al.(2021){Evoli}, {Amato}, {Blasi}, and {Aloisio}]{Evoli2021prdb}
{Evoli}, C.; {Amato}, E.; {Blasi}, P.; {Aloisio}, R.
\newblock {Stochastic nature of Galactic cosmic-ray sources}.
\newblock {\em \prd} {\bf 2021}, {\em 104},~123029,  \href{http://arxiv.org/abs/2111.01171}{{\normalfont [arXiv:astro-ph.HE/2111.01171]}}.
\newblock {\url{https://doi.org/10.1103/PhysRevD.104.123029}}.

\bibitem[{Adriani} et~al.(2009){Adriani}, {Barbarino}, {Bazilevskaya}, {Bellotti}, {Boezio}, {Bogomolov}, {Bonechi}, {Bongi}, {Bonvicini}, {Bottai}, {Bruno}, {Cafagna}, {Campana}, {Carlson}, {Casolino}, {Castellini}, {de Pascale}, {de Rosa}, {de Simone}, {di Felice}, {Galper}, {Grishantseva}, {Hofverberg}, {Koldashov}, {Krutkov}, {Kvashnin}, {Leonov}, {Malvezzi}, {Marcelli}, {Menn}, {Mikhailov}, {Mocchiutti}, {Orsi}, {Osteria}, {Papini}, {Pearce}, {Picozza}, {Ricci}, {Ricciarini}, {Simon}, {Sparvoli}, {Spillantini}, {Stozhkov}, {Vacchi}, {Vannuccini}, {Vasilyev}, {Voronov}, {Yurkin}, {Zampa}, {Zampa}, and {Zverev}]{positrons.PAMELA}
{Adriani}, O.; {Barbarino}, G.C.; {Bazilevskaya}, G.A.; {Bellotti}, R.; {Boezio}, M.; {Bogomolov}, E.A.; {Bonechi}, L.; {Bongi}, M.; {Bonvicini}, V.; {Bottai}, S.;  et~al.
\newblock {An anomalous positron abundance in cosmic rays with energies 1.5-100GeV}.
\newblock {\em \nat} {\bf 2009}, {\em 458},~607--609,  \href{http://arxiv.org/abs/0810.4995}{{\normalfont [arXiv:astro-ph/0810.4995]}}.
\newblock {\url{https://doi.org/10.1038/nature07942}}.

\bibitem[{Amato} and {Blasi}(2018)]{Amato2018asr}
{Amato}, E.; {Blasi}, P.
\newblock {Cosmic ray transport in the Galaxy: A review}.
\newblock {\em Advances in Space Research} {\bf 2018}, {\em 62},~2731--2749,  \href{http://arxiv.org/abs/1704.05696}{{\normalfont [arXiv:astro-ph.HE/1704.05696]}}.
\newblock {\url{https://doi.org/10.1016/j.asr.2017.04.019}}.

\bibitem[{Orusa} et~al.(2022){Orusa}, {Di Mauro}, {Donato}, and {Korsmeier}]{Orusa2022prd}
{Orusa}, L.; {Di Mauro}, M.; {Donato}, F.; {Korsmeier}, M.
\newblock {New determination of the production cross section for secondary positrons and electrons in the Galaxy}.
\newblock {\em \prd} {\bf 2022}, {\em 105},~123021,  \href{http://arxiv.org/abs/2203.13143}{{\normalfont [arXiv:astro-ph.HE/2203.13143]}}.
\newblock {\url{https://doi.org/10.1103/PhysRevD.105.123021}}.

\bibitem[{Cavasonza} et~al.(2017){Cavasonza}, {Gast}, {Kr{\"a}mer}, {Pellen}, and {Schael}]{Cavasonza2017apj}
{Cavasonza}, L.A.; {Gast}, H.; {Kr{\"a}mer}, M.; {Pellen}, M.; {Schael}, S.
\newblock {Constraints on Leptophilic Dark Matter from the AMS-02 Experiment}.
\newblock {\em \apj} {\bf 2017}, {\em 839},~36,  \href{http://arxiv.org/abs/1612.06634}{{\normalfont [arXiv:hep-ph/1612.06634]}}.
\newblock {\url{https://doi.org/10.3847/1538-4357/aa624d}}.

\bibitem[{Cholis} and {McKinnon}(2022)]{Cholis2022prd}
{Cholis}, I.; {McKinnon}, I.
\newblock {Constraining the charge-, time-, and rigidity-dependence of cosmic-ray solar modulation with AMS-02 observations during Solar Cycle 24}.
\newblock {\em \prd} {\bf 2022}, {\em 106},~063021,  \href{http://arxiv.org/abs/2207.12447}{{\normalfont [arXiv:astro-ph.SR/2207.12447]}}.
\newblock {\url{https://doi.org/10.1103/PhysRevD.106.063021}}.

\bibitem[{Ptuskin} et~al.(2013){Ptuskin}, {Zirakashvili}, and {Seo}]{Ptuskin2013apj}
{Ptuskin}, V.; {Zirakashvili}, V.; {Seo}, E.S.
\newblock {Spectra of Cosmic-Ray Protons and Helium Produced in Supernova Remnants}.
\newblock {\em \apj} {\bf 2013}, {\em 763},~47,  \href{http://arxiv.org/abs/1212.0381}{{\normalfont [arXiv:astro-ph.HE/1212.0381]}}.
\newblock {\url{https://doi.org/10.1088/0004-637X/763/1/47}}.

\bibitem[{Recchia} and {Gabici}(2018)]{Recchia2018mnras}
{Recchia}, S.; {Gabici}, S.
\newblock {Non-linear acceleration at supernova remnant shocks and the hardening in the cosmic ray spectrum}.
\newblock {\em \mnras} {\bf 2018}, {\em 474},~L42--L46,  \href{http://arxiv.org/abs/1710.01111}{{\normalfont [arXiv:astro-ph.HE/1710.01111]}}.
\newblock {\url{https://doi.org/10.1093/mnrasl/slx191}}.

\bibitem[{Yuan} et~al.(2011){Yuan}, {Zhang}, and {Bi}]{Yuan2011prd}
{Yuan}, Q.; {Zhang}, B.; {Bi}, X.J.
\newblock {Cosmic ray spectral hardening due to dispersion in the source injection spectra}.
\newblock {\em \prd} {\bf 2011}, {\em 84},~043002,  \href{http://arxiv.org/abs/1104.3357}{{\normalfont [arXiv:astro-ph.HE/1104.3357]}}.
\newblock {\url{https://doi.org/10.1103/PhysRevD.84.043002}}.

\bibitem[{Maurin} et~al.(2023){Maurin}, {Ahlers}, {Dembinski}, {Haungs}, {Mangeard}, {Melot}, {Mertsch}, {Wochele}, and {Wochele}]{Maurin2023crdb}
{Maurin}, D.; {Ahlers}, M.; {Dembinski}, H.; {Haungs}, A.; {Mangeard}, P.S.; {Melot}, F.; {Mertsch}, P.; {Wochele}, D.; {Wochele}, J.
\newblock {A cosmic-ray database update: CRDB v4.1}.
\newblock {\em European Physical Journal C} {\bf 2023}, {\em 83},~971,  \href{http://arxiv.org/abs/2306.08901}{{\normalfont [arXiv:astro-ph.HE/2306.08901]}}.
\newblock {\url{https://doi.org/10.1140/epjc/s10052-023-12092-8}}.

\end{thebibliography}

\end{document}